# The Effect of Dielectric Capping on Few-Layer Phosphorene Transistors: Tuning the Schottky Barrier Heights

Han Liu, Adam T. Neal, Mengwei Si, Yuchen Du and Peide D. Ye, *Fellow, IEEE*

*Abstract* – **Phosphorene is a unique single elemental semiconductor with two-dimensional layered structures. In this letter, we study the transistor behavior on mechanically exfoliated few-layer phosphorene with the top-gate. We achieve a high on-current of 144 mA/mm and hole mobility of 95.6 cm$^2$/V·s. We deposit Al$_2$O$_3$ by atomic layer deposition (ALD) and study the effects of dielectric capping. We observe that the polarity of the transistors alternated from p-type to ambipolar with Al$_2$O$_3$ grown on the top. We attribute this transition to the changes for the effective Schottky barrier heights for both electrons and holes at the metal contact edges, which is originated from fixed charges in the ALD dielectric.**

## I. INTRODUCTION

THE rise of graphene has brought great opportunities for two-dimensional (2D) materials. MoS$_2$, a semiconducting 2D crystal, has been given intensive research in the past years as it shows interesting electrical and optical properties and has great potential in future electronics and optoelectronic devices. [1-5] However, due to S-vacancy defect level or charge neutral level of MoS$_2$ located at the vicinity of conduction band edge, MoS$_2$ transistors are mostly showing n-type behavior. [4-8] This places a serious constraint to CMOS applications, where a p-type transistor is needed as a complementary component to reduce static power consumption in logic circuits and systems. [8]

In this letter, we introduce transistors based on a novel 2D narrow band gap semiconductor, few-layers of black phosphorus or phosphorene, as a channel material for p-type transistors. The black phosphorus, first obtained back in 1914, is the least reactive one in all allotropes of phosphorus at room temperature. [9] Away from the polymeric structure of the red phosphorus, the phosphorene is an orthogonal crystal with layered structures. Each layer with thickness of ~0.6 nm, consisted of hexagonal honeycombs, is very similar to graphene. However, it differs from graphene that all honeycombs are three-fold puckered along one direction so that it has a bulk band gap of ~0.3 eV due to the broken symmetry. Few-layer p-type phosphorene transistors were demonstrated very recently. [10-15]

This material is based upon work partly supported by NSF under Grant CMMI-1120577 and SRC under Tasks 2362 and 2396.
Han Liu, Adam T. Neal, Mengwei Si, Yuchen Du and Peide D. Ye are with the School of Electrical and Computer Engineering and the Birck Nanotechnology Center, Purdue University, West Lafayette, IN 47907 USA (Tel: +1 (765) 494-7611, Fax: +1 (765) 496-7443, E-mail: yep@purdue.edu).

In this study, we will focus on ALD capping effect on high-performance p-type few-layer phosphorene transistors. Positive fixed charges in dielectric efficiently tune the effective Schottky barrier heights at metal/phosphorene contact edge and change the transistors' polarity from p-type to ambipolar. This shows a promising technique to realize threshold voltage (V$_T$) adjustment and hence alternate the transistor polarity on 2D semiconductors.

## II. EXPERIMENT

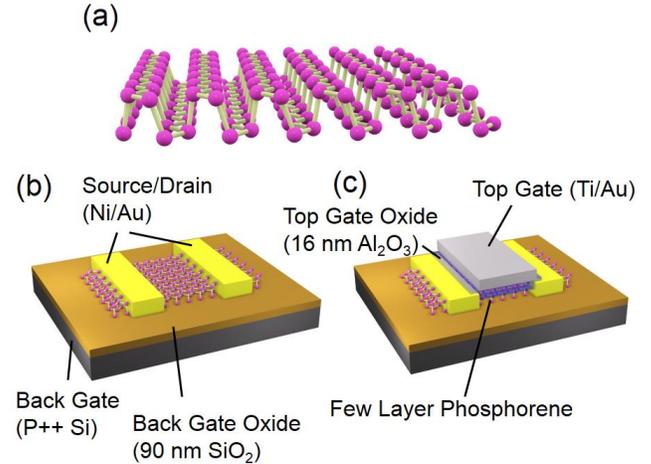

**Fig. 1** (a) Crystal structure of a single layer phosphorene. (b) Schematic view of a back-gate few-layer phosphorene p-type transistor. Ni/Au are used as source/drain contacts. (c) Schematic view of a double-gate few-layer phosphorene transistor. A 16 nm Al$_2$O$_3$ is used as gate dielectric and Ti/Au is used as metal gate for an ambipolar transistor.

The crystal structure of a single layer phosphorene is shown in Fig. 1(a). We start from mechanical exfoliation from commercially available bulk crystals. By standard scotch-tape techniques, multi-layers of phosphorene of thickness around 4-8 nm were transferred to heavily doped silicon substrate which was capped with 90 nm SiO$_2$. We found out flake thickness below this range would have a low on-current and over this range would have a significantly dropped current on/off ratio. [11] Source/drain regions were defined by e-beam lithography, followed by a Ni/Au metallization of 20/60 nm with e-beam evaporator. The gap between source and drain was 500 nm. After the device was first measured with back gate only, the top gate dielectric was grown by ALD. In order to secure a uniform dielectric layer, a 0.8 nm Al was deposited prior to ALD process. The Al layer was oxidized in atmosphere and served as a seeding layer. Afterwards, a 15 nm $Al_2O_3$ was deposited with trimethylaluminum (TMA) and $H_2O$ as precursors at 250 °C. The total dielectric thickness would be ~16 nm if we count the oxidized Al on the thickness of layer. Top gate was defined with e-beam lithography and Ti/Au of 20/60 nm was used as the top gate. Gate length was designed as same as channel length ($L_g=L_{ch}=500$ nm). Final device structures are depicted in Fig. 1(b) and (c). All measurements were carried with Keithley 4200 Semiconductor Characterization System at room temperature.

### III. RESULTS AND DISCUSSION

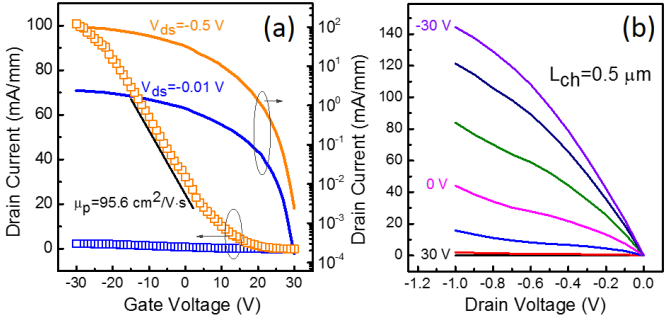

**Fig.2** (a) Transfer characteristics of a back gate phosphorene transistor. Both log scale (left axis) and linear scale (right axis) of the drain current are shown. (b) Output characteristics of the same transistor.

We first measure the device performance prior to dielectric capping. The transfer characteristics are shown in Fig. 2(a) on a 6 nm thick device. Drain voltage is biased at -0.01 and -0.5 V. We observe a clear p-type transistor behavior. A current on/off ratio over $10^4$ is observed, showing a good switching behavior for this narrow band gap 2D semiconductor. The maximum transconductance ($g_m$) for -0.01 and -0.5 V bias is 0.071 and 3.6 mS/mm. By using $g_m=\mu_p C_{ox} V_{ds} W/L$, where $C_{ox}$ is the oxide capacitance, $V_{ds}$ is the drain bias, and $W/L$ is the width/length of the channel, we can have an estimation of the field-effect hole mobility $\mu_p$ of 95.6 $cm^2/V\cdot s$. The intrinsic value of the field effective hole mobility in few-layer phosphorene transistors may further surpass this calculated value, as 1) the large contact resistance due to the existence of a Schottky barrier at the interface, 2) the high density of oxide traps in the thick dielectric layer, and 3) the anisotropic mobility distribution in the 2D plane of phosphorene.[11,13] The output characteristics can be seen in Fig. 2(b). The back gate bias is applied from 30 to -30 V with a 10 V step, while the drain bias swept from 0 to -1 V. A linear current-voltage relationship at low $V_{ds}$ shows a relatively good contact property at metal/phosphorene interface, however, a Schottky barrier for holes does exist at the interface, as will be discussed later. A maximum drain current of 144 mA/mm is achieved at -1 V drain bias. This is larger than the drain current in most of other 2D transistors, due to the narrow band gap and high mobility of phosphorene.[10-12] Previous studies have shown that substrates and dielectric would have a strong impact on transport properties in 2D semiconductors, as they generate remote phonons as well as screen the extrinsic charge impurities.

For any transistor technology, a top-gate process must be developed as well as a passivation technique for the 2D surface.

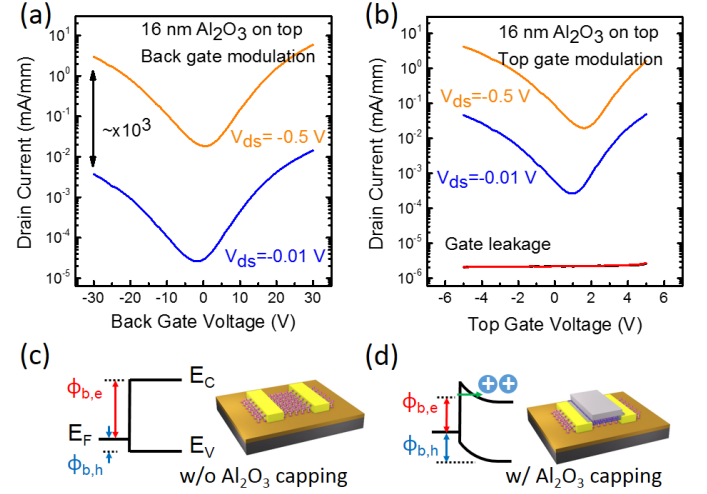

**Fig.3** (a)(b)Back gate and top gate transfer curves of phosphorene transistor after dielectric capping. (c)(d) Band diagrams showing effective Shottky barrier heights for electrons/holes at the contact edges.

We further study the effect of dielectric capping on device performance. Fig. 3(a) and 3(b) show the transfer curves from both back gate and top gate modulations after ALD $Al_2O_3$ capping. The drain biases are maintained at -0.01 and -0.5 V, the same as the measurements shown in Fig. 2. Low leakage current level down to $10^{-6}$ mA/mm is observed for top-gate, showing good dielectric quality. Different from a clear p-type transistor behavior, it shows an ambipolar behavior, where both hole current and electron current are observed. For back gate modulated transfer curves, a maximum hole current of 3 mA/mm at -30 V gate bias is achieved at -0.5 V drain bias. Compared to a 100 mA/mm hole current prior to $Al_2O_3$ capping, the drain current drops by 30 times. Meanwhile, we observe an electron current of 5.9 mA/mm at the other side of the curve. If we take a look at the top-gate modulation in Fig. 3(b), similar curves are observed. This transition from p-type to ambipolar is rarely observed in other transistors based on transition metal dichalcogenides. We attribute this transition to the reduction of effective Schottky heights for electrons at the contact edge, due to the modulation of surface charges by ALD $Al_2O_3$. In conventional MOSFETs on bulk semiconductors, the polarity of the transistor, either n-type or p-type, is defined by the source/drain dopants. However, for transistors on semiconducting 2D crystals, in the absence of source/drain implantation, the transistor polarity is more likely to be dominated by the band alignment at the metal contact. Previous studies on metal contacts on $MoS_2$ have revealed that metals are pinned at the vicinity of conduction band edge, making $MoS_2$ transistor mostly n-type.[6,8] Similarly, the p-type behavior without dielectric capping of phosphorene would indicate the Fermi-level is pinned near the valence band, as depicted in Fig. 3(c). Assuming a flat-band condition here, therefore, the heights of Schottky barriers for electrons and holes are simply defined by $\Phi_{b,e}=E_C-E_F$ and $\Phi_{b,h}=E_F-E_V$, where $\Phi_{b,e}$ and $\Phi_{b,h}$ are the barrier heights for electrons and holes, $E_C$, $E_F$, $E_V$ are the energy levels of conduction band edge, contact metal, and valence band edge. In this case, $\Phi_{b,h}$ is much smaller than $\Phi_{b,e}$, thus phosphorene transistor behaves like p-type prior to dielectric capping. However, with $Al_2O_3$ capping, positive fixed charges in the dielectric layer would induce a band bending. The positive fixed charges are originated from lower ALD growth temperature. [16] Therefore, the effective barrier heights for electrons and holes would be changed. This effect is expected to be more obvious for narrow band gap materials. As shown in Fig. 3(d), due to the band bending, the effective barrier height for electrons is reduced, thus it would be easier for electrons injected from the metal to the conduction band through a thermal-assisted tunneling process. Meanwhile, barrier heights for holes are enlarged, reducing the hole current. Once the barrier heights for electrons and holes are comparable at zero gate bias, the device is more likely to behave as an ambipolar transistor, leaving large Schottky barriers for both electrons and holes. This can be identified from the transfer curve in Fig. 3(a), where the ratio of on-current for -0.01 V and -0.5 V drain bias is over $10^3$, showing a typical barrier transport behavior. In the meantime, comparing Fig. 3(a) to Fig. 2(a), an increased subthreshold slope and hence a much higher off-state current was observed. This may indicate the increase of interface trap density, due to the degradation of phosphorene top surface during ALD process. Previous studies have revealed that $H_2O$ molecules, one of the precursors in our ALD $Al_2O_3$ process, can serve as a surface oxidant for phosphorene under certain circumstances. However, this surface degradation, or the enhanced interface trap density, does not affect the polarity of the transistor, as the contact regions are protected by the metals and remain intact during ALD growth. But still, this advices us the necessity of the optimization for the ALD process for better device performance.

Understanding the effects of dielectric capping is helpful for better device designs for 2D semiconductors. On one hand, the band edge alignment at metal/semiconductor interface determines the transport behavior of the transistors. In great contrast to conventional MOSFETs, where the charges in dielectric or interface traps change the carrier density in the channel, they tune the effective barrier heights in transistors on 2D semiconductors. Since the drain current is more sensitive to barrier heights than carrier density, the dielectric capping would have a more significant impact on transistors with 2D crystals. By designing and optimizing the dielectric growth process such as growth temperature and post deposition annealing, we can effectively adjust the threshold voltage of the transistors, rather than changing the work function of the gate metal.

## IV. CONCLUSION

In summary, we have demonstrated well behaved transistors based on phosphorene ultrathin films. An on-current of 144 mA/mm and hole mobility of 95.6 cm$^2$/V·s have been achieved. After capping ALD $Al_2O_3$ gate dielectric, we observe a transition from p-type to ambipolar transistor behavior. We attribute this change to positive fixed charges in dielectric which alternates the Schottky barrier heights for electrons and holes. This opens a new way in $V_T$ adjustment for 2D transistors and possible realization of 2D CMOS circuits by dielectric engineering.